\documentclass[11pt]{article}
\pdfoutput=1 
\usepackage{amsmath,amssymb,amsbsy,amstext, amsthm, simplewick, amsfonts}
\usepackage{graphicx}
\usepackage{epsfig}
\usepackage{verbatim} 
\usepackage{fancybox}
\usepackage{color}
\usepackage{mathdots}
\usepackage{ulem}
\usepackage{enumitem}
\usepackage{caption}
\usepackage{mathtools}
\usepackage{youngtab}
\usepackage{bbm}
\usepackage{parskip}
\usepackage{booktabs}
\usepackage[numbers,sort&compress]{natbib}
\usepackage{ytableau}
\usepackage[utf8]{inputenc}
\usepackage{tikz}
\usetikzlibrary{positioning, trees,decorations, decorations.markings, decorations.pathmorphing, arrows, shapes.geometric}

\usepackage{tocloft}

\allowdisplaybreaks

\newcommand{\Comment}[1]{{}}
\definecolor{darkblue}{rgb}{0.15,0.35,0.55}
\definecolor{reddish}{rgb}{0.65, 0.2, 0.2}
\definecolor{darkgreen}{RGB}{50,150,0}
\usepackage[linktocpage=true]{hyperref}
\hypersetup{
colorlinks=true,
citecolor=darkblue,
linkcolor=reddish,
urlcolor=darkblue,
pdfauthor={},
pdftitle={},
pdfsubject={}
}

\usepackage{cleveref}

\newcommand{\be}{\begin{equation}}
\newcommand{\ee}{\end{equation}}

\newcommand{\rd}{{\rm d}}

\numberwithin{equation}{section}

\makeatletter
\def\thickhline{%
  \noalign{\ifnum0=`}\fi\hrule \@height \thickarrayrulewidth \futurelet
   \reserved@a\@xthickhline}
\def\@xthickhline{\ifx\reserved@a\thickhline
               \vskip\doublerulesep
               \vskip-\thickarrayrulewidth
             \fi
      \ifnum0=`{\fi}}
\makeatother

\newlength{\thickarrayrulewidth}
\begin{document}

\renewcommand{\thefootnote}{\fnsymbol{footnote}}
~
\vspace{0truecm}
\thispagestyle{empty}
\begin{center}
{\fontsize{20}{18} \bf Exceptional scalar theories in de Sitter space}\\[8pt] 
\end{center} 

\vspace{.7truecm}

\begin{center}
\hspace{-.3cm} {\fontsize{13.5}{18}\selectfont
James Bonifacio,${}^{\rm a}$\footnote{\href{mailto:jb2389@cam.ac.uk}{\texttt{jb2389@cam.ac.uk}}} Kurt Hinterbichler,${}^{\rm b}$\footnote{\href{mailto:kurt.hinterbichler@case.edu}{\texttt{kurt.hinterbichler@case.edu}}} Austin Joyce,${}^{\rm c}$\footnote{\href{mailto:austinjoyce@uchicago.edu}{\texttt{austinjoyce@uchicago.edu}}} and Diederik Roest${}^{\rm d}$\footnote{\href{mailto:d.roest@rug.nl}{\texttt{d.roest@rug.nl}}}
}
\end{center}
\vspace{.4truecm}

 \centerline{{\it ${}^{\rm a}$Department of Applied Mathematics and Theoretical Physics,}}
 \centerline{{\it Cambridge University, Cambridge, CB3 0WA, UK}} 
 
   \vspace{.3cm}

 \centerline{{\it ${}^{\rm b}$CERCA, Department of Physics,}}
 \centerline{{\it Case Western Reserve University, 10900 Euclid Ave, Cleveland, OH 44106, USA}} 
 
  \vspace{.3cm}
 
  \centerline{{\it ${}^{\rm c}$Kavli Institute for Cosmological Physics, Department of Astronomy and Astrophysics,}}
 \centerline{{\it University of Chicago, Chicago, IL 60637, USA}}
 
  \vspace{.3cm}

 \centerline{{\it ${}^{\rm d}$Van Swinderen Institute for Particle Physics and Gravity, University of Groningen,}}
 \centerline{{\it Nijenborgh 4, 9747 AG Groningen, The Netherlands}}
 

\vspace{.6cm}
\begin{abstract}
\noindent
The special galileon and Dirac--Born--Infeld (DBI) theories are effective field theories of a single scalar field that have many interesting properties in flat space. These theories can be extended to all maximally symmetric spaces, where their algebras of shift symmetries are simple. We study aspects of the curved space versions of these theories: for the special galileon, we find a new compact expression for its Lagrangian in de Sitter space and a field redefinition that relates it to the previous, more complicated  formulation. This field redefinition reduces to the well-studied galileon duality redefinition in the flat space limit.
For the DBI theory in de Sitter space, we discuss the brane and dilaton formulations of the theory and present strong evidence that these are related by a field redefinition.  We also give an interpretation of the symmetries of these theories in terms of broken diffeomorphisms of de Sitter space. 

\end{abstract}

\newpage

\setcounter{tocdepth}{2}
\tableofcontents
\newpage

\renewcommand*{\thefootnote}{\arabic{footnote}}
\setcounter{footnote}{0}

\section{Introduction}

Self-interactions for massless spinning particles are severely constrained; with minimal assumptions, the only possibilities are Yang--Mills theory for spin-1 particles and general relativity for spin-2 particles~\cite{Weinberg:1965nx,Benincasa:2007xk,Schuster:2008nh,McGady:2013sga}.  However, without additional assumptions such broad statements cannot be made for scalar field theories, where Lorentz symmetry is less constraining.
One powerful additional assumption is that a theory has a nonlinearly realized symmetry. This leads to certain exceptional effective field theories, where nonlinearly realized symmetries fix interaction terms with special properties. The on-shell avatar of a nonlinearly realized symmetry is a soft theorem for amplitudes. In the simplest case, this implies that an amplitude has a zero in the soft limit, i.e.,  the amplitude scales with a positive power of an external momentum as this momentum is sent to zero. 

Examples of exceptional theories include nonlinear sigma models and the Dirac--Born--Infeld (DBI) scalar field theory. A nonlinear sigma model has a nonlinearly realized internal symmetry and amplitudes that scale linearly with the soft momentum, i.e., an Adler zero \cite{Adler:1964um} (see Ref.~\cite{Cheung:2021yog} for a recent geometric interpretation of soft theorems), while the DBI theory has a nonlinearly realized higher-dimensional Poincar\'e symmetry and a quadratic scaling of amplitudes in the soft limit. 
Beyond these two examples, there is a single possibility for a single scalar field with a larger nonlinear spacetime symmetry and a cubic scaling of amplitudes in the soft limit, which is called the special galileon.\footnote{The flat space special galileon is a particular example of the more general scalar galileon theories \cite{Nicolis:2008in}, which have shift symmetries that are constant or linear in the spacetime coordinates. The special galileon has an additional shift symmetry that is quadratic in the coordinates \cite{Hinterbichler:2015pqa}.} No larger spacetime symmetries can be nonlinearly realized and no non-trivial higher scaling can be obtained in the soft limit \cite{ Cheung:2014dqa, Bogers:2018zeg, Roest:2019oiw}, given certain assumptions.
In addition to these theories with vanishing soft limits, there is a scalar theory where the soft limit obeys a nontrivial soft theorem, namely the dilaton, which spontaneously breaks conformal symmetry down to Poincar\'e symmetry
\cite{Callan:1970yg,Boels:2015pta,Huang:2015sla,DiVecchia:2015jaq,DiVecchia:2017uqn}.  This dilaton theory can also be realized as the DBI theory of a brane embedded into an AdS space of one dimension higher \cite{Goon:2011qf,Goon:2011uw}.
The two formulations are related by a complicated field redefinition involving all powers of the field \cite{Bellucci:2002ji, Creminelli:2013fxa}.

Aside from their improved soft behavior, these exceptional scalar theories display other interesting structures. For example, they possess an intricate web of relationships---including relations to theories of massless particles~\cite{Cachazo:2014xea,Cheung:2017ems,Cheung:2017yef}---and interesting double copy structures~\cite{Bern:2019prr}. Additionally, these scalar theories can be thought of as analogues of gravity in a precise sense~\cite{Sundrum:2003yt,Klein:2015iud,Bonifacio:2019rpv}. The rich interconnections are indicative of recurring structural motifs in quantum field theory that the study of these theories can help to uncover.

The above paradigm is well-established for flat space. There are good reasons to investigate possible generalizations for the other maximally symmetric spaces. Chief amongst these are cosmology and holography: in the former, the inflationary period is usually modeled as close to de Sitter (dS) space while, in the latter, one considers gravity duals that asymptote to anti-de Sitter (AdS) space.   
In this paper, we will for definiteness write things in the language of dS space in Lorentzian signature, though everything we consider can be straightforwardly extended to AdS spaces and to any signature.  

The possible shift symmetries of a scalar field in dS space can be conveniently classified using flat ambient space. For $D$-dimensional dS space with coordinates $x^{\mu}$, $\mu=0, \dots, D-1$, we specify its embedding into the ambient space $\mathbb{R}^{D,1}$ through the ambient space coordinates $X^A(x)$, $A=1, \dots, D+1$, that satisfy $X^2(x)\equiv \eta_{AB} X^A(x) X^B(x)= H^{-2}$, where the ambient space metric is $\eta_{AB}={\rm diag}(-1, 1, \dots, 1)$. 
A dS scalar field $\phi(x)$ can be represented by an ambient space field $\Phi(X)$ satisfying $\Phi(X(x))=\phi(x) $ and $\Phi(\lambda X) = \lambda^w \Phi(X)$ for $\lambda>0$, where the weight $w$ specifies how to continue $\Phi$ away from the dS surface (see Ref.~\cite{Bonifacio:2018zex} for more details of the ambient space formalism in our conventions).  The possible shift symmetries of $\Phi$ can then be written as
\be \label{eq:general-shift}
\delta \Phi = S_{A_1 \dots A_k} X^{A_1} \dots X^{A_k} (X^2 H^2)^{(w-k)/2} + \dots ,
\ee
where $\dots$ denotes possible field-dependent terms, $S_{A_1 \dots A_k}$ is a constant, symmetric, traceless tensor in ambient space, and $k$ is an integer that denotes the order of the shift symmetry. Restricting the transformation \eqref{eq:general-shift} to the dS surface gives the shift symmetry of $\phi$, which is independent of $w$. For a given $k$, the mass of $\phi$ consistent with the shift symmetry is given by
\be
m^2_k = -k(k+D-1)H^2.
\ee
Although these squared masses are negative in dS space, the shift-symmetric scalars correspond to unitary exceptional series representations of the dS group \cite{Basile:2016aen,Sun:2021thf}. Their Euclidean sphere partition functions were calculated in Ref.~\cite{Law:2020cpj} using techniques developed in Ref.~\cite{Anninos:2020hfj}.  

It is a non-trivial problem to deform these shift symmetries to non-commuting symmetries and find interactions that are invariant under the resulting symmetry algebras they form with the dS isometries.
In Ref.~\cite{Bonifacio:2018zex}, a manifestly $\mathbb{Z}_2$-invariant interacting Lagrangian for the case $k=2$ was found.   This theory is the dS version of the special galileon. The full nonlinearly realized symmetry algebra is $\mathfrak{sl}(D+1, \mathbb{R})$, which is spontaneously broken to the dS isometries.  The expression for this Lagrangian in general dimensions takes a rather complicated form involving hypergeometric functions.  In this paper, we employ a kind of duality transformation---a field redefinition that reduces to galileon duality in the flat limit---to find another presentation of the theory.  This new presentation is far simpler and has an interpretation in terms of broken diffeomorphisms, at the expense of obscuring the $\mathbb{Z}_2$ symmetry.

For $k=1$, there naively appears to be two interacting theories on dS space that realize an algebra with non-commuting shift symmetries.  One is the dS DBI theory, constructed by embedding a dS brane into an AdS space of one dimension higher \cite{Goon:2011qf,Goon:2011uw}.  Another is the dS conformal dilaton \cite{Hinterbichler:2012mv}, which nonlinearly realizes conformal symmetry on dS space.
These are the dS versions of the brane and dilaton realizations of conformal symmetry in flat space.  In both cases, the full algebra is $\frak{so}(D+1,1)$, which is broken to the dS isometries.   We will argue that these two theories are equivalent, by finding a perturbative field redefinition that relates the two theories up to very high order in the fields.  This is the dS version of the transformation of Ref.~\cite{Bellucci:2002ji} and is consistent with our expectation that the symmetry breaking pattern alone determines the nonlinear theory governing the interactions of the Goldstones (this has been proven for internal symmetries \cite{Coleman:1969sm}, but not for spacetime symmetries).

Finally, we will see that the transformations for all of these nonlinear shift-symmetric theories on dS space can be given a geometric interpretation.  They can be seen as arising from a subset of the infinitesimal diffeomorphisms of dS space.  The diffeomorphisms of dS space include the Killing vectors which leave dS invariant, as well as certain ``exact diffeomorphisms" whose infinitesimal vector fields can be written as the gradients of scalars.   Subsets of these exact diffeomorphisms can be identified with the nonlinearly realized symmetries realized by the scalar fields, which are the Goldstone modes for the spontaneous breaking down to the Killing symmetries.

\paragraph{Conventions:}
We denote the spacetime dimension by $D$.  We use the mostly plus metric signature convention and the curvature conventions of Ref.~\cite{Carroll:2004st}.    We denote the dS space Hubble scale as $H$, so that the Ricci scalar is $R=D(D-1)H^2>0$. Tensors are symmetrized and antisymmetrized with unit weight, e.g., $T_{(\mu\nu)}={1\over 2} \left(T_{\mu\nu}+T_{\nu\mu}\right)$ and $T_{[\mu\nu]}={1\over 2} \left(T_{\mu\nu}-T_{\nu\mu}\right)$. We define $\epsilon_{01\cdots D} = 1$.

\section{Special galileon in dS space}

We start with the $k=2$ theory, which is the special galileon in dS space. A rather complicated Lagrangian for this theory was written in Ref.~\cite{Bonifacio:2018zex}, with the advantage that the ${\mathbb Z}_2$ symmetry $\phi\rightarrow-\phi$ of the model is manifest.  Here we find a much simpler Lagrangian that is not manifestly ${\mathbb Z}_2$ symmetric.  We then show how these two Lagrangians are mapped into one another by a field redefinition that is like a dS version of galileon duality for the highest-derivative terms.   
Galileon duality in flat space \cite{deRham:2013hsa} is a type of invertible field redefinition that preserves the galileon-like structure of the Lagrangian---it is local order-by-order in powers of the field, but it involves an infinite series of terms.

\subsection{Symmetry algebra}

In the ambient space $\mathbb{R}^{D,1}$, the generators of the dS isometries are packaged into an antisymmetric tensor, $J_{AB}$, that satisfies the $\mathfrak{so}(D,1)$ commutation relations, 
\be \left[ J_{A_1 A_2},J_{B_1 B_2}\right]= \eta_{A_1 B_1}J_{A_2 B_2}-\eta_{A_2 B_1}J_{A_1 B_2}+\eta_{A_2 B_2}J_{A_1 B_1}-\eta_{A_1 B_2}J_{A_2 B_1} \,.
\label{eq:Jcomm}
\ee
These isometries are realized on an ambient space scalar field as
\be 
\delta_{J_{AB}}\Phi \equiv J_{AB}\Phi=X_A\partial_B\Phi-X_B\partial_A\Phi \,.
\ee
The generators of the shift symmetries for a $k=2$ scalar are packaged into a symmetric, traceless tensor, $S_{AB}$, which transforms as a tensor under the isometries,
\be
[J_{A_1 A_2}, S_{B_1 B_2}] =  \eta_{A_1 B_1} S_{A_2 B_2} - \eta_{A_2 B_1} S_{A_1 B_2}+ \eta_{A_1 B_2} S_{A_2 B_1} -\eta_{A_2 B_2} S_{A_1 B_1}.
\label{k2commutatoralge}
\ee
The unique possibility for the remaining commutators can be written as \cite{Bonifacio:2018zex}
\be
[S_{A_1 A_2}, S_{B_1 B_2}] = -\frac{\alpha^2}{\Lambda^{D+2}} \left( \eta_{A_1 B_1} J_{A_2 B_2} + \eta_{A_2 B_1} J_{A_1 B_2}+ \eta_{A_1 B_2} J_{A_2 B_1} + \eta_{A_2 B_2} J_{A_1 B_1} \right),
\label{k2commutatoralge}
\ee
where $\alpha$ is a dimensionless constant and $\Lambda>0$ is an energy scale.
For $\alpha^2>0$, these commutation relations correspond to the algebra $\mathfrak{sl}(D+1, \mathbb{R})$; this is true for any spacetime signature or sign of the curvature. For $\alpha^2<0$, the commutation relations correspond to a real form of $\mathfrak{sl}(D+1, \mathbb{C})$ that depends on the spacetime signature and the sign of the curvature. When $\alpha^2 = 0$, we get the undeformed algebra of a free theory. 

\subsection{Manifestly $\mathbb{Z}_2$-invariant Lagrangian}

We start by reviewing the formulation of the dS galileon given in Ref.~\cite{Bonifacio:2018zex}. We nonlinearly realize the algebra $\mathfrak{sl}(D+1, \mathbb{R})$ on a scalar field through the following ambient space transformation:
\be
\label{eq:delta-sgal-Z2}
\delta \Phi = S_{AB} \left(X^{A} X^B-\frac{\alpha^2}{\Lambda^{D+2}} \partial^{A}\Phi\partial^{B}\Phi \right), 
\ee
where the ambient space field has weight two,\footnote{We can always change the weight by rescaling the field by powers of $X^2H^2$. This will modify the ambient space form of the symmetry transformation but does not affect its form in the physical dS space.}
\be
X^A \partial_A \Phi = 2\Phi.
\ee
The ambient space transformation induces a transformation on the dS space field $\phi(x) = \Phi(X(x))$ that we can write as\footnote{To find this, we replace $S_{AB}={1\over 2}\partial_A\partial_B(S_{CD}X^CX^D)$ and then use the embedding space reduction rules discussed in Ref.~\cite{Bekaert:2010hk}.} 
\be
\delta \phi=\sigma-\frac{\alpha^2}{\Lambda^{D+2}} \left( {1\over 2} \nabla_{\mu}\nabla_{\nu}\sigma \nabla^\mu\phi\nabla^\nu\phi +H^2 \sigma(\nabla\phi)^2+2H^2\phi \nabla_\mu\phi\nabla^\mu \sigma+4H^4 \sigma\phi^2 \right),
\ee
where $\sigma \equiv S_{AB} X^{A}(x)X^{B}(x)$.

A Lagrangian invariant under the above transformation was found in Ref.~\cite{Bonifacio:2018zex} and describes the unique ghost-free interacting theory of a $k=2$ scalar, which is the dS space version of the special galileon.  In terms of the dimensionless field $\hat{\phi} \equiv - 2\alpha H^2 \phi/ \Lambda^{(D+2)/2}$, this Lagrangian is 
\begin{align} 
\label{eq:dS-sgal-Z2}
\frac{{\cal L}}{\sqrt{-g}}=& \frac{\Lambda^{D+2}}{4H^2 \alpha^2 } \Bigg[
\sum_{j=0}^{D-1}{(1+\hat{\phi})^{D-j}+ (-1)^{j}(1-\hat{\phi})^{D-j} \over (2H^2)^{(j+1)}(1-\hat{\phi}^2)^{\frac{D+3}{2}}\,\Gamma(j+3)} \left[(j+1)f_{j+1}(\hat{\phi})-(j+2) f_j(\hat{\phi}) \right] \partial^\mu\hat{\phi}\partial^\nu \hat{\phi} X^{(j)}_{\mu\nu}(\hat{\Pi}) \nonumber \\
& -\frac{2}{ (D+1)}\left(1-{(1+\hat{\phi})^{D+1}+(1-\hat{\phi})^{D+1}\over 2 (1-\hat{\phi}^2)^{(D+1)/2}}\right) \Bigg],
\end{align}
where we have defined
\be
f_j (\hat{\phi})\equiv  {}_2F_1\left({D+3\over 2},{j+1\over 2};{j+3\over 2};{(\partial \hat{\phi})^2\over 4H^2(1-\hat{\phi}^2)} \right)\,,
\ee
and where $X_{\mu \nu}^{(j)}$ are defined recursively by
\be
X^{(0)}_{\mu\nu}(\hat{\Pi})  \equiv g_{\mu \nu}, \qquad\quad X^{(n)}_{\mu \nu}(\hat{\Pi}) \equiv -n \hat{\Pi}_{\mu}{}^{\alpha} X_{\alpha \nu}^{(n-1)}(\hat{\Pi})+g_{\mu \nu}\hat{\Pi}^{\alpha \beta} X^{(n-1)}_{\alpha \beta}(\hat{\Pi}),
\ee
with $\hat{\Pi}_{\mu \nu} \equiv \nabla_{\mu} \nabla_{\nu} \hat{\phi}$. 
When expanded out, this Lagrangian contains only terms with even powers of the field, so it is manifestly invariant under the $\mathbb{Z}_2$ symmetry that acts as $\phi \rightarrow - \phi$. 

\subsection{A simplified special galileon Lagrangian}

There are other ways to nonlinearly realize the algebra $\mathfrak{sl}(D+1, \mathbb{R})$ on a scalar field. Motivated by the reasoning described in Section \ref{sec:broken-diffs}, we consider the following ambient space transformation:
\be \label{eq:sgal-new-ambient}
\delta \Phi = S_{AB} \left( \frac{1}{H^2X^2}X^{A} X^{B}+\frac{2\alpha}{\Lambda^{(D+2)/2}} X^{A }\partial^{B} \Phi \right) ,
\ee
where the ambient space field has weight zero,
\be
X^A \partial_A \Phi = 0.
\ee
In terms of the dS field $\phi(x) = \Phi(X(x))$, this transformation is 
\be \label{eq:sgal-new-dS}
\delta  \phi=\sigma+\frac{\alpha}{\Lambda^{(D+2)/2}}\nabla_\mu\phi\nabla^\mu \sigma,
\ee
where $\sigma \equiv S_{AB} X^{A}(x)X^{B}(x)$.
This transformation satisfies the same commutation relations and symmetry breaking pattern as the transformation \eqref{eq:delta-sgal-Z2} and, as we will see, leads to a much simpler formulation of the dS space special galileon.

We now search for a ghost-free action invariant under the transformation \eqref{eq:sgal-new-dS}. By imposing invariance under the symmetry order by order in the field and then resumming, we get the following result:
\be
\label{eq:Lsgal-new}
\frac{\mathcal{L}}{\sqrt{-g}} = \frac{\Lambda^{D+2}}{4\alpha^2 H^2} \left[ \frac{2}{D+1} \left( \cosh (D+1) \hat{\phi} -1 \right)- e^{-(D+1) \hat{\phi}} \sum_{j=0}^{D-1} (-1)^j\frac{\partial^{\mu} \hat{\phi} \partial^{\nu} \hat{\phi} X^{(j)}_{\mu \nu}(\hat{\Pi})}{(j+2)!H^{2j+2}}\right].
\ee
This is much simpler than the manifestly $\mathbb{Z}_2$-invariant Lagrangian \eqref{eq:dS-sgal-Z2}, although the $\mathbb{Z}_2$ symmetry is no longer manifest.\footnote{Note also that with this Lagrangian we cannot realize other real forms of $\mathfrak{sl}(D+1, \mathbb{C})$ on real fields.}
The equations of motion also take a relatively simple form,
\be
\label{eq:sgalEOM}
\frac{2\alpha}{\Lambda^{(D+2)/2} } \frac{\delta \mathcal{L}}{\delta \phi}  =-e^{(D+1) \hat{\phi}}-\frac{e^{-(D+1)\hat{\phi}}}{D!} \epsilon^{\mu_1 \dots \mu_D} \epsilon^{\nu_1\dots  \nu_D} G_{\mu_1 \nu_1} \dots G_{\mu_D \nu_D} =0\,,
\ee
where 
\be
G_{\mu \nu} \equiv g_{\mu \nu}-\frac{1}{H^2} \nabla_{\mu} \nabla_{\nu} \hat{\phi}+\frac{1}{H^2} \partial_{\mu} \hat{\phi} \partial_{\nu} \hat{\phi}.
\label{eq:k2metric}
\ee
This tensor $G_{\mu \nu}$ transforms covariantly, like a metric, under the shift symmetry \eqref{eq:sgal-new-dS}. We can use it to build higher-order invariants and to couple to other matter fields, as in Ref.~\cite{Bonifacio:2019rpv}.

\subsection{Mapping between Lagrangians}

We now have two formulations of the special galileon Lagrangian in dS space: the manifestly $\mathbb{Z}_2$-invariant Lagrangian \eqref{eq:dS-sgal-Z2} from Ref.~\cite{Bonifacio:2018zex} and the new simplified Lagrangian \eqref{eq:sgal-new-dS}, which is invariant under a shift symmetry that is linear in the field.   Since these realize the same symmetry breaking pattern and are both ghost-free theories of the same derivative order, we expect that these two Lagrangians are related to each other by a field redefinition.  We now show that this is indeed the case and that the field redefinition can be thought of as a dS uplift of the flat space galileon duality transformations described in Refs.~\cite{Fasiello:2013woa,deRham:2013hsa}.

We start from the ambient space transformation in Eq.~\eqref{eq:delta-sgal-Z2},
\be \label{eq:delta-sgal-Z2-2}
\delta \Phi^{(0)} = S_{AB} \left( X^{A} X^{B}-\frac{\alpha^2}{\Lambda^{D+2}} \partial^{A}\Phi^{(0)} \partial^{B}\Phi^{(0)} \right),
\ee
where $\Phi^{(0)}$ has weight two. This describes the shift symmetry of the manifestly $\mathbb{Z}_2$-invariant Lagrangian in Eq.~\eqref{eq:dS-sgal-Z2}.
Next we make a galileon duality transformation in its active form \cite{Kampf:2014rka} on the ambient space field,
\be \label{eq:duality-ambient}
{\Phi}^{(1)}= e^{-\theta \bar{\delta}}\Phi^{(0)}, \quad {\rm with}\quad \bar{\delta} \Phi^{(0)} \equiv-  \frac{\alpha}{2\Lambda^{(D+2)/2}}\partial_{A} \Phi^{(0)} \partial^{A} \Phi^{(0)}, 
\ee
where $\theta$ is a dimensionless real parameter. 
Under this field redefinition, the ambient transformation in Eq.~\eqref{eq:delta-sgal-Z2-2} becomes \cite{Hinterbichler:2015pqa}
\be \label{eq:delta-sgal-mixed}
\delta \Phi^{(1)} =  S_{AB} \left(X^A X^B+ \frac{ 2\alpha\theta }{\Lambda^{(D+2)/2}} X^A \partial^B  \Phi^{(1)}+ \frac{\alpha^2(\theta^2-1 )}{\Lambda^{D+2}} \partial^A  \Phi^{(1)} \partial^B \Phi^{(1)} \right).
\ee
We now set $\theta = 1$, so we get
\be
\delta \Phi^{(1)} =  S_{AB} \left(X^A X^B+\frac{2\alpha}{\Lambda^{(D+2)/2}} X^A \partial^B  \Phi^{(1)} \right),
\ee
with only a linear term in $\Phi$. 
Now define the weight zero field $\Phi^{(2)} \equiv \Phi^{(1)}/H^2 X^2$, so that
\be
\delta \Phi^{(2)} =  S_{AB} \left(\frac{X^A X^B}{H^2 X^2}\left(1+ \frac{4 \alpha H^2 \Phi^{(2)}}{\Lambda^{(D+2)/2}} \right)+\frac{2\alpha}{\Lambda^{(D+2)/2}} X^A \partial^B  \Phi^{(2)} \right).
\ee
Finally, we define the weight zero field
\be
\Phi^{(3)} \equiv \frac{\Lambda^{(D+2)/2}}{4 \alpha H^2} \log \left(1+ \frac{4 \alpha H^2 \Phi^{(2)}}{\Lambda^{(D+2)/2}} \right),
\ee
so that we get
\be
\delta \Phi^{(3)} =  S_{AB} \left(\frac{X^A X^B}{H^2 X^2}+\frac{2\alpha}{\Lambda^{(D+2)/2}} X^A \partial^B  \Phi^{(3)} \right),
\ee
which is the form of the shift symmetry of the new simplified Lagrangian \eqref{eq:sgal-new-dS}.

In summary, we have a redefinition defined by
\be
\hat{\Phi}^{(3)} = -\frac{1}{2 } \log \left(1- \frac{2  }{X^2 H^2 } e^{-\theta \bar{\delta}} \hat{\Phi}^{(0)}\right), \quad\qquad \bar{\delta} \hat{\Phi}^{(0)} = \frac{1}{4H^2} (\partial \hat{\Phi}^{(0)})^2, 
\ee
where $\theta =1$ and $\hat{\Phi}^{(i)} \equiv -2\alpha H^2 \Phi^{(i)}/\Lambda^{(D+2)/2}$. In terms of physical dS fields, this is
\be \label{eq:final-redef-ambient}
\hat{\phi}^{(3)} = -\frac{1}{2 } \log \left(1- 2  e^{-\theta \bar{\delta}}\hat{\phi}^{(0)}\right), \quad \bar{\delta} \hat{\phi}^{(0)} = {1\over 4 H^2} \left( (\partial \hat{\phi}^{(0)}  )^2+4 H^2 (\hat{\phi}^{(0)})^2 \right),
\ee
where $\theta =1$ and $\hat{\phi}^{(i)} \equiv -2\alpha H^2 \phi^{(i)}/\Lambda^{(D+2)/2}$.
As a check of this result, if we just consider the zero-derivative part of the transformation, then the field redefinition is\footnote{Defining $\bar{\delta}_0 \hat{\phi}=\hat{\phi}^2$ gives $\bar{\delta}_0^m \hat{\phi}^n=\frac{(n+m-1)!}{(n-1)!} \hat{\phi}^{n+m}$, so then we get 
\be
-\frac{1}{2} \log\left(1-2 e^{-\bar{\delta}_0}\hat{\phi}\right) = \sum_{n=1}^{\infty} \sum_{m=0}^{\infty} \frac{2^{n-1}}{n}\frac{(-1)^m}{m!} \frac{(n+m-1)!}{(n-1)!} \hat{\phi}^{n+m} = \tanh^{-1} \hat{\phi}.
\ee}
\be
\hat{\phi}^{(3)} = \tanh^{-1}\hat{\phi}^{(0)},
\ee
under which we can check that the potential terms of the two Lagrangians get mapped into each other,
\be
\cosh \left( (D+1) \hat{\phi}^{(3)} \right) = {\left(1+\hat{\phi}^{(0)}\right)^{D+1}+\left(1-\hat{\phi}^{(0)}\right)^{D+1}\over 2 \left(1-(\hat{\phi}^{(0)})^2\right)^{\frac{D+1}{2}}} .
\ee
For general $\theta$, the transformation \eqref{eq:final-redef-ambient} leads to a one-parameter family of Lagrangians related by the ambient space duality.  In the flat space limit $H\rightarrow 0$, this becomes the flat space galileon duality transformation \cite{Fasiello:2013woa,deRham:2013hsa} in its active form \cite{Kampf:2014rka} acting on the special galileon.

\section{DBI and conformal dilaton in dS space}

In this section, we discuss two formulations of the nonlinear $k=1$ theory in dS space, corresponding to the brane and dilaton realizations of conformal symmetry.  Given that they both describe the same symmetry breaking pattern, we expect them to be related by a field redefinition.  We show up to some high order in the fields that this is indeed the case.

\subsection{Symmetry algebra}

The shift symmetries for a $k=1$ scalar are packaged into a $(D+1)$-dimensional ambient space vector $S_{A}$.  Since they transform as a vector under the isometries, we have the commutator
\be
[J_{A B}, S_{C}] =   \eta_{AC} S_{B } - \eta_{BC} S_{A} .
\label{k2commutatoralge}
\ee
The unique commutator which completes the algebra can be written as
\be
[S_{A}, S_{B}] = - \frac{\alpha^2H^2}{ \Lambda^{D+2}}  J_{AB}.
\label{k2commutatoralge}
\ee
For $\alpha^2>0$, these commutation relations correspond to the conformal algebra algebra $\mathfrak{so}(D,2)$. For $\alpha^2< 0$ the algebra is $\mathfrak{so}(D+1,1)$, while for AdS$_D$ it is $\mathfrak{so}(D-1,3)$ if $\alpha^2>0$ or $\mathfrak{so}(D,2)$ if $\alpha^2<0$.

\subsection{DBI Lagrangian}

One way to nonlinearly realize the conformal algebra $\mathfrak{so}(D,2)$ on a scalar field in dS space is through the ambient space transformation
\be
\delta \Phi = S_A \left( X^A - \frac{\alpha^2H^2}{ \Lambda^{D+2}} \Phi \partial^A \Phi \right),\label{confdbiinsfle}
\ee
where $\Phi$ has weight $w=1$.
A set of interactions invariant under this symmetry were given in general dimensions in Ref.~\cite{Bonifacio:2018zex}, but they were quite complicated since the interaction ansatz used there did not allow for their simplest form. If instead we look for a Lagrangian in the $P(\phi,(\partial\phi)^2)$ 
form, then there is a simple invariant interaction without a tadpole,
\be \label{eq:DBI-sqrt}
\frac{\mathcal{L}}{\sqrt{-g}}  =  \frac{ \Lambda^{D+2}}{\alpha^2 H^2}\frac{1}{(1- \hat{\phi}^2)^{D/2}} \sqrt{1- \frac{  (\partial \hat{\phi})^2/H^2}{1- \hat{\phi}^2}}\,,
\ee
where now $\hat{\phi} \equiv - \alpha H^2 \phi/ \Lambda^{(D+2)/2}$. Defining $ \hat{\pi} \equiv \tanh^{-1}\hat{\phi}$, this Lagrangian can be written as
\be \label{eq:dbi-cosh}
\frac{\mathcal{L}}{\sqrt{-g}} =  \frac{ \Lambda^{D+2}}{\alpha^2 H^2}\cosh^D \hat{\pi}\sqrt{1- \frac{  (\partial \hat{\pi})^2}{H^2 \cosh^2 \hat{\pi}}}.
\ee
The tadpole interaction can be written as
\be
\frac{\mathcal{L}'}{\sqrt{-g}} =- \frac{\Lambda^{D+2}\hat{\phi} }{\alpha^2 H^2} \, _2F_1\left(\frac{1}{2},\frac{D+2}{2} ;\frac{3}{2}; \hat{\phi}^2 \right).
\ee
These Lagrangians realize the symmetry breaking pattern $\mathfrak{so}(D,2) \rightarrow \mathfrak{so}(D,1)$ and can be realized physically as the DBI theory of a dS$_D$ brane probing an AdS$_{D+1}$ bulk.\footnote{Defining instead $ \hat{\pi} \equiv \tanh^{-1}(1/\hat{\phi})$, the Lagrangian can be written as
\be \label{eq:dbi-sinh}
\frac{\mathcal{L}}{\sqrt{-g}} =  -\frac{ \Lambda^{D+2}}{\alpha^2 H^2}i^D \sinh^D \hat{\pi}\sqrt{1+ \frac{  (\partial \hat{\pi})^2}{H^2 \sinh^2 \hat{\pi}}}\,,
\ee
which for $D=4$ takes the same form as the DBI theory of a dS$_4$ brane in an AdS$_5$ bulk of Refs.~\cite{Goon:2011qf,Goon:2011uw} (see also Ref.~\cite{Grall:2019qof}).
In the case of AdS, i.e., replacing $H \rightarrow i/L$ in Eq.~\eqref{eq:dbi-cosh}, we get the theory of an AdS$_D$ brane in an AdS$_{D+1}$ bulk from Ref.~\cite{Clark:2005ht}, which realizes the symmetry breaking pattern $\mathfrak{so}(D,2) \rightarrow \mathfrak{so}(D-1,2)$.}

From the dS DBI Lagrangian \eqref{eq:DBI-sqrt}, we can take various limits to recover other known theories.
For example, taking the flat limit $H\rightarrow 0$ and $\Lambda \rightarrow 0$ with $\Lambda_{\rm flat} \equiv  (\Lambda^{D+2}/H^2)^{1/D}$ and $\phi$ held fixed, we recover the theory of a flat brane in a two-time Minkowski bulk spacetime,
\be
\lim_{H \rightarrow 0}\frac{\mathcal{L}}{\sqrt{-g}} =\frac{\Lambda_{\rm flat} ^{D}}{\alpha^2} \sqrt{1- \frac{\alpha^2 (\partial \phi)^2}{ \Lambda_{\rm flat} ^{D}}}\,,
\ee
i.e., flat space DBI with the sign of the quartic interaction that violates positivity constraints \cite{Adams:2006sv}.\footnote{Starting instead from AdS$_D$ with $\Lambda_{\rm flat} \equiv  (\Lambda^{D+2} L^2)^{1/D}$  fixed gives the standard DBI theory,
\be
\lim_{L \rightarrow \infty}\frac{\mathcal{L}}{\sqrt{-g}}  =-\frac{\Lambda_{\rm flat} ^{D}}{\alpha^2} \sqrt{1+ \frac{\alpha^2 (\partial \phi)^2}{ \Lambda_{\rm flat} ^{D}}}\,,
\ee
corresponding to $\mathfrak{iso}(D,1) \rightarrow \mathfrak{iso}(D-1,1)$.} This corresponds to the symmetry breaking pattern $\mathfrak{iso}(D-1,2) \rightarrow \mathfrak{iso}(D-1,1)$. If we instead take the limit $\Lambda\rightarrow0$ with $H$ fixed, then Eq.~\eqref{eq:DBI-sqrt} reduces after rescaling to the theory of a dS brane probing a flat background \cite{Goon:2011qf,Goon:2011uw}, 
\be
\lim_{\Lambda \rightarrow 0}\frac{H^{(D+2)^2/2}}{\Lambda^{(D+2)^2/2}} \frac{\mathcal{L}}{\sqrt{-g}} =\frac{H^{D^2/2}}{i^D\alpha^{D+2} } {1\over \phi^D} \sqrt{1+ { (\partial \phi)^2\over H^2 \phi^2} },
\ee
corresponding to $\mathfrak{iso}(D,1) \rightarrow \mathfrak{so}(D,1)$.\footnote{The same symmetry breaking pattern is realized by the dS galileon \cite{Goon:2011qf,Goon:2011uw,Burrage:2011bt}.} The final limit we consider is more subtle since we must give the field a vacuum expectation value that we send to $- \infty$. We replace $\hat{\pi} \rightarrow \ln(H/ 2\Lambda)+ \hat{\pi} $ in Eq.~\eqref{eq:dbi-cosh} and then take the limit $H \rightarrow 0$ with $\Lambda$ and $\hat{\pi}$ held fixed. Adding the tadpole interaction and rescaling the Lagrangian, this gives the theory describing a flat brane in a two-time dS bulk spacetime,
\be \label{eq:2-time-dS}
\lim_{H \rightarrow 0} \frac{H^{D+2}}{\Lambda^{D+2}} \frac{1}{\sqrt{-g}}\left( \mathcal{L} -D  \mathcal{L}' \right)= \frac{\Lambda^{D}}{\alpha^2 }e^{-D \hat{\pi} } \left(\sqrt{1-  e^{2 \hat{\pi}  }\frac{(\partial \hat{\pi})^2 }{\Lambda^2}}-1\right).
\ee
This corresponds to the symmetry breaking pattern $\mathfrak{so}(D,2) \rightarrow \mathfrak{iso}(D-1,1)$.\footnote{If we instead start in AdS$_D$ and replace $\hat{\pi} \rightarrow -\ln( 2\Lambda L)+ \hat{\pi} $, then we  get
\be \label{eq:conformal-dbi}
\lim_{L \rightarrow \infty} \frac{1}{(\Lambda L)^{D+2}}\frac{1}{\sqrt{-g}} \left( \mathcal{L} -D \mathcal{L}' \right) = -\frac{\Lambda^{D}}{\alpha^2 }e^{-D \hat{\pi} } \left(\sqrt{1+  e^{2 \hat{\pi}  }\frac{(\partial \hat{\pi})^2 }{\Lambda^2}}-1\right).
\ee
This is the conformal DBI theory, which describes a flat brane in an AdS bulk spacetime \cite{deRham:2010eu}. This realizes the same symmetry breaking pattern $\mathfrak{so}(D,2) \rightarrow \mathfrak{iso}(D-1,1)$ as the Lagrangian in Eq.~\eqref{eq:2-time-dS}. While these two Lagrangians are inequivalent, as can be seen by calculating their $2\rightarrow 2$ scattering amplitudes, we expect there to be a field redefinition between them once we include the other invariant interactions. }

\subsection{Conformal dilaton Lagrangian}

An alternative way to nonlinearly realize the algebra $\mathfrak{so}(D,2)$ on a scalar field in dS space is through the following ambient space transformation:
\be \label{eq:delta-dS-conf-dilaton}
\delta \Phi = S_A \left( \frac{1}{\sqrt{H^2 X^2}} X^A +\frac{\alpha}{\Lambda^{(D+2)/2}} \sqrt{H^2 X^2}\partial^A \Phi \right),
\ee
where $\Phi$ has weight $w=0$. This is the form of the shift symmetry suggested by the discussion of Section~\ref{sec:broken-diffs}.   It satisfies the same commutation relations as the transformation in Eq.~\eqref{confdbiinsfle} and realizes the same symmetry breaking pattern $\mathfrak{so}(D,2) \rightarrow \mathfrak{so}(D,1)$.

If we look for invariant interactions of the $P(\phi,(\partial\phi)^2)$  form, then for the interaction without a tadpole we get
\be \label{eq:k=1-dilaton}
\frac{\mathcal{L}}{\sqrt{-g}} =  \frac{\Lambda^{D+2}}{2\alpha^2H^2} \left[ -\frac{1}{H^2}e^{(D-2)  \hat{\phi}} (\partial \hat{\phi})^2  +e^{D \hat{\phi}}-\frac{De^{(D-2)  \hat{\phi}} -2 }{(D-2)}\right],
\ee
where again $\hat{\phi} \equiv - H^2 \alpha \phi/ \Lambda^{(D+2)/2}$. This agrees with the interaction in Ref.~\cite{Hinterbichler:2012mv} when $D=4$. For $D=2$, the last term in square brackets should be understood as the $D \rightarrow 2$ limit.  When $D=1$, the conformal dilaton Lagrangian \eqref{eq:k=1-dilaton} coincides with the special galileon Lagrangian \eqref{eq:Lsgal-new} in a space with half the Hubble constant, $ H_{k=2}= H_{k=1}/2$. This is possible due to the Lie algebra isomorphism $\mathfrak{so}(2,1) = \mathfrak{sl}(2, \mathbb{R})$. 

There are additional independent invariant Lagrangians. For example, in every dimension there is the invariant tadpole term $e^{D \hat{\phi} }$.
In any even dimension, there is also a special interaction from the point of view of cohomology, the Wess--Zumino term, which in dS$_4$ is \cite{Hinterbichler:2012mv}
\begin{align}
\frac{\mathcal{L}'}{\sqrt{-g}} &= -\frac{ \Lambda^6}{\alpha^2 H^2} \left[ \hat{\phi} -\frac{1}{4 }e^{4  \hat{\phi}} +\frac{1}{2 H^2} (\partial \hat{\phi})^2-\frac{1}{6H^4 } (\partial \hat{\phi})^2 \Box \hat{\phi} -\frac{1}{12 H^4  }(\partial \hat{\phi})^4  \right].
\end{align}
There is also an interaction with the potential $\cosh^D( \hat{\phi} )$, which maps to the square root DBI action \eqref{eq:DBI-sqrt} under the field redefinition discussed below. In dS$_4$, this Lagrangian is
\be
\begin{aligned} 
 \frac{\mathcal{L}}{\sqrt{-g}} =\,& -\frac{ \Lambda^6}{\alpha^2 H^2} \bigg[\frac{1}{2H^2}
      (\partial \hat{\phi})^2 e^{- \hat{\phi} } \cosh ^2( \hat{\phi}) (\cosh \hat{\phi} -3 \sinh  \hat{\phi} )-\cosh ^4( \hat{\phi} )   \\
   &+\frac{11}{48H^8}  (\partial \hat{\phi})^8 e^{-4  \hat{\phi} }+\frac{1}{24H^6}  (\partial \hat{\phi})^6 e^{-4  \hat{\phi}} \left(e^{2  \hat{\phi} }-4\right)-\frac{1}{24H^4}  (\partial \hat{\phi})^4 e^{-4  \hat{\phi}} \left(2 e^{2  \hat{\phi}}+1\right)   \\
      & -\frac{1}{H^2} \mathcal{L}^{\rm TD}_1(\hat{\Pi}) \left(\frac{5}{12H^6} (\partial \hat{\phi})^6 e^{-4  \hat{\phi}  }+\frac{1}{24H^4}   (\partial \hat{\phi})^4 e^{-4  \hat{\phi} } \left(e^{2  \hat{\phi} }+1\right)+\frac{1}{2H^2}   (\partial \hat{\phi})^2 e^{-3  \hat{\phi} } \cosh  \hat{\phi} \right)  \\
   &+ \frac{1}{H^4}  \mathcal{L}^{\rm TD}_2(\hat{\Pi}) \left(\frac{1}{4H^4}   (\partial \hat{\phi})^4 e^{-4  \hat{\phi} }+\frac{1}{12H^2}   (\partial \hat{\phi})^2 e^{-4  \hat{\phi}
   } \left(e^{2  \hat{\phi} }+2\right)\right) \\
   &-\frac{1}{12 H^8}  \mathcal{L}^{\rm TD}_3(\hat{\Pi})  (\partial \hat{\phi})^2  e^{-4  \hat{\phi} } \bigg],
\end{aligned}
\ee
where $\mathcal{L}^{\rm TD}_n$ are terms that are total derivatives in the flat space limit and are defined recursively by
\be
\mathcal{L}^{\rm TD}_n( \hat{\Pi}) =  \sum_{j=1}^n (-i)^{j+1} \frac{(n-1)!}{(n-j)!} [ \hat{\Pi}^j] \mathcal{L}^{\rm TD}_{n-j}( \hat{\Pi}),
\ee
with $[ \hat{\Pi}^j] \equiv \Pi_{\mu_1}{}^{\mu_2}  \Pi_{\mu_2}{}^{\mu_3} \dots \Pi_{\mu_j}{}^{\mu_1}$ and $\mathcal{L}^{\rm TD}_0( \hat{\Pi}) =1$. 

As with the DBI Lagrangian \eqref{eq:DBI-sqrt}, we can take limits of the dilaton Lagrangian \eqref{eq:k=1-dilaton} to recover other theories. Taking $H, \Lambda \rightarrow 0$ with $H^2/\Lambda^{(D+2)/2}$ fixed gives the kinetic term of the conformal galileon,
\be
\lim_{H, \Lambda \rightarrow 0} \frac{\mathcal{L}}{\sqrt{-g}} = -\frac{1}{2} e^{(D-2)  \hat{\phi}} (\partial \phi)^2,
\ee
which corresponds to the symmetry breaking pattern $\mathfrak{so}(D,2) \rightarrow \mathfrak{iso}(D-1,1)$. This corresponds to the same symmetry breaking pattern as the theory of a flat brane in two-time dS space in Eq.~\eqref{eq:2-time-dS} and conformal DBI in Eq.~\eqref{eq:conformal-dbi}. A field redefinition that maps linear combinations of the conformal galileons into linear combinations of the conformal DBI interactions was worked out in Refs.~\cite{Bellucci:2002ji, Creminelli:2013fxa}, which shows the equivalence of these flat space theories. 

\subsection{Mapping between Lagrangians}
We can demonstrate the equivalence of the dS DBI and conformal dilaton Lagrangians by finding a field redefinition that maps between them, as with the two formulations of the dS special galileon. The transformation maps each dS conformal dilaton Lagrangian into a certain linear combination of the dS DBI Lagrangians,  as with the flat space versions of these theories \cite{Bellucci:2002ji,Creminelli:2013fxa}. For example, as mentioned above, the combination of conformal dilaton Lagrangians with potential  $\cosh^D( \hat{\phi} )$ maps to the square root DBI action \eqref{eq:DBI-sqrt}.

We start with the DBI transformation
\be
\delta \Phi^{(0)} = S_A \left( X^A - \frac{\alpha^2H^2}{ \Lambda^{D+2}} \Phi^{(0)}  \partial^A \Phi^{(0)}  \right),
\label{eq:k1deltaP0}
\ee
where $\Phi^{(0)} $ has weight one. Now set $\Phi^{(1)} = e^{- \theta \bar{\delta}} \Phi^{(0)}$, where $\bar{\delta}$ is defined such that for $\theta=1$ we have
\be \label{eq:k=1-delta-Phi1}
\delta \Phi^{(1)} =S_A \left( X^A + \frac{\alpha}{ \Lambda^{(D+2)/2}} \sqrt{H^2 X^2}  \partial^A \Phi^{(1)}  \right).
\ee
Defining $\hat{\Phi}^{(0)} \equiv - H^2 \alpha \Phi^{(0)}/ \Lambda^{(D+2)/2}$, we find by a brute force calculation that, at least up to high orders, we can write the action of $\bar{\delta}$ perturbatively in ambient space as
\begin{align} 
\bar{\delta} \hat{\Phi}^{(0)} &\equiv \sqrt{X^2H^2} \frac{(\partial \hat{\Phi}^{(0)})^{2}}{H^{2}} \left(\frac{1}{2}-\frac{1}{4} \frac{\hat{\Phi}^{(0)}}{\sqrt{X^2H^2}}-\frac{1}{8}\frac{(\hat{\Phi}^{(0)})^2}{X^2H^2}+\dots \right) +\sqrt{X^2H^2}\frac{(\partial \hat{\Phi}^{(0)})^{4}}{H^{4}} \left( \frac{1}{16}+\dots \right)+\dots \nonumber \\
& =  -\sqrt{X^2H^2}\sum_{j=1}^{\infty} \frac{(\partial \hat{\Phi}^{(0)})^{2j}}{H^{2j}} f_j\left( \frac{\hat{\Phi}^{(0)}}{\sqrt{X^2H^2}} \right), \label{eq:k=1-derivation}
\end{align}
where $f_j$ is a power series. We do not know a closed-form expression for the $f_j$, but we have computed the terms contributing to  $\bar{\delta}$ up to 17th order in the fields, which are listed in Table~\ref{tab:functions}.

\bgroup
\def\arraystretch{1.25}%
\begin{table}[!h]
  \centering
{\scriptsize
 \begin{tabular}{ l  }
\hspace{6cm} 
Expansions of $f_j(x)$\\ \hline\rule{0pt}{7ex}    
$
\begin{aligned}
 f_1(x)=&-\frac{1}{2}+\frac{x}{4}+\frac{x^2}{8}+\frac{x^3}{24}+\frac{x^4}{96}+\frac{x^5}{80}+\frac{7 x^6}{480}+\frac{19
   x^7}{6720}-\frac{37 x^8}{17920}+\frac{197 x^9}{13440}+\frac{113
   x^{10}}{8960}-\frac{41323 x^{11}}{1182720}-\frac{53429 x^{12}}{14192640} \nonumber \\
   &+\frac{14633519 x^{13}}{92252160}-\frac{1095239 {x
   }^{14}}{12300288}-\frac{285950669 x^{15}}{369008640} + \dots ,
\end{aligned}$\\
\rule{0pt}{7ex}    
$
\begin{aligned}
f_2(x)=&-\frac{1}{16}+\frac{x^2}{32}+\frac{x^3}{64}-\frac{x^4}{120}+\frac{13 x^5}{3840}+\frac{101
   x^6}{4480}-\frac{281 x^7}{17920}-\frac{1093 x^8}{26880}+\frac{40387 x^9}{430080}+\frac{206869
   x^{10}}{2365440}-\frac{4802311 x^{11}}{9461760} \nonumber \\
   &-\frac{222421 x^{12}}{20500480}+\frac{923059867
   x^{13}}{295206912}+ \dots,
   \end{aligned}$\\
\rule{0pt}{7ex}    
$
\begin{aligned}
f_3(x) =& -\frac{1}{64}-\frac{x}{96}+\frac{11 x^2}{960}+\frac{23 x^3}{1920}-\frac{209
   x^4}{13440}+\frac{59 x^5}{53760}+\frac{1523 x^6}{26880}-\frac{979 x^7}{15360}-\frac{7505
   x^8}{39424}+\frac{773923 x^9}{1351680} \nonumber \\
   & +\frac{104962307 x^{10}}{184504320}-\frac{317494109
   x^{11}}{67092480}+\dots ,
   \end{aligned}$\\
\rule{0pt}{7ex}    
$
\begin{aligned}   
   f_4(x) =&-\frac{17}{3840}-\frac{61 x}{7680}+\frac{149 x^2}{35840}+\frac{411 x^3}{35840}-\frac{10481
   x^4}{430080}-\frac{17 x^5}{86016}+\frac{286673 x^6}{1892352}-\frac{2239247
   x^7}{9461760}-\frac{809516299 x^8}{984023040}\nonumber \\
   &+\frac{1380091937 x^9}{421724160} + \dots,\end{aligned}$\\
\rule{0pt}{5ex}    
$
\begin{aligned}   
  f_5(x) =& -\frac{19}{17920}-\frac{179 x}{35840}+\frac{137 x^2}{107520}+\frac{2677
   x^3}{215040}-\frac{99019 x^4}{2365440}+\frac{24859 x^5}{9461760}+\frac{54077893
   x^6}{123002880}-\frac{28892971 x^7}{30750720} + \dots, \end{aligned}$\\
\rule{0pt}{5ex}    
$
\begin{aligned}   
   f_6(x) =&-\frac{13}{430080}-\frac{417 x}{143360}+\frac{677 x^2}{946176}+\frac{58859
   x^3}{3784704}-\frac{6993213 x^4}{82001920}+\frac{27430969 x^5}{984023040} + \dots,\end{aligned}$\\
\rule{0pt}{5ex}    
$
\begin{aligned}   
   f_7(x) = &\frac{189}{901120}-\frac{32833 x}{18923520}+\frac{205627 x^2}{98402304}+\frac{34589413
   x^3}{1476034560}+ \dots,\end{aligned}$\\
\rule{0pt}{5ex}    
$
\begin{aligned}   
   f_8(x) =&\frac{52727}{281149440}-\frac{558697 x}{393609216}+ \dots\end{aligned}$
  \end{tabular}
  }
  \caption{\small Power series expansion of the first few $f_j$s appearing in Eqs.~\eqref{eq:k=1-derivation} and~\eqref{eq:eqwithfjs}. These are the contributions to~\eqref{eq:k=1-derivation} up to 17th order in fields. These expressions can be extended to higher orders with additional effort, though we have not been able to find closed-form formulas for them.
  } \label{tab:functions}
\end{table}
\egroup

For general values of $\theta$, the first few terms in the shift transformation of $\Phi^{(1)} = e^{- \theta \bar{\delta}} \Phi^{(0)}$ are
\be
\begin{aligned}
\delta \Phi^{(1)} =S_A &\bigg(X^A + \frac{ \theta \alpha}{ \Lambda^{(D+2)/2}} \sqrt{H^2 X^2}  \partial^A \Phi^{(1)} + \frac{H^2\alpha^2(\theta-1)}{4\Lambda^{D+2}} \left[  2(\theta+2)  \Phi^{(1)}  \partial^A \Phi^{(1)}  - \theta X^A \left(\partial \Phi^{(1)}\right)^2 \right]  \\
& + \frac{H^4\alpha^3 \theta (\theta-1)}{4\Lambda^{3(D+2)/2} \sqrt{H^2 X^2}} \left[  \left( X^2 \partial^A \Phi^{(1)}  -X^A \Phi^{(1)}\right)  \left(\partial \Phi^{(1)}\right)^2+ \partial^A \Phi^{(1)} \left(\Phi^{(1)}\right)^2 \right] + \dots \bigg), \label{eq:k=1-delta-Phi1-theta}
\end{aligned}
\ee
so the shift transformation truncates to linear or quadratic order in the fields only for $\theta \in \{0, 1\}$. Note that we use the homogeneity condition $X^A \partial_A \Phi^{(1)} =\Phi^{(1)}$ to get Eq.~\eqref{eq:k=1-delta-Phi1-theta}.

Setting $\theta=1$, we now define the weight zero field $\Phi^{(2)} \equiv \Phi^{(1)}/\sqrt{H^2 X^2}$, so that
\be
\delta \Phi^{(2)} =  S_{A} \left(\frac{X^A}{\sqrt{H^2 X^2}}\left(1+ \frac{ \alpha H^2 \Phi^{(2)}}{\Lambda^{(D+2)/2}} \right)+\frac{\alpha \sqrt{H^2 X^2}}{\Lambda^{(D+2)/2}} \partial^A  \Phi^{(2)} \right).
\ee
Finally, we define the weight zero field
\be
\Phi^{(3)} \equiv \frac{\Lambda^{(D+2)/2}}{ \alpha H^2} \log \left(1+ \frac{ \alpha H^2 \Phi^{(2)}}{\Lambda^{(D+2)/2}} \right),
\ee
so that we get
\be
\delta \Phi^{(3)} =  S_A \left( \frac{1}{\sqrt{H^2 X^2}} X^A +\frac{\alpha}{\Lambda^{(D+2)/2}} \sqrt{H^2 X^2}\partial^A \Phi^{(3)} \right),
\ee
which is the dS conformal dilaton transformation \eqref{eq:delta-dS-conf-dilaton}.

To summarize, the fields in the brane and dilaton presentations of the symmetry breaking pattern $\mathfrak{so}(D,2) \rightarrow \mathfrak{so}(D,1)$ are related by the ambient space field redefinition
\be
\hat{\Phi}^{(3)} = - \log \left(1-  \frac{e^{-  \bar{\delta}} \hat{\Phi}^{(0)}}{\sqrt{H^2 X^2}}\right),
\ee
where $\hat{\Phi}^{(i)} \equiv -\alpha H^2 \Phi^{(i)}/\Lambda^{(D+2)/2}$ and the derivation $\bar{\delta}$ is defined in Eq.~\eqref{eq:k=1-derivation}. In terms of dS fields, this is
\be
\hat{\phi}^{(3)} = - \log \left(1-e^{- \bar{\delta}} \hat{\phi}^{(0)}\right),  \quad \bar{\delta} \hat{\phi}^{(0)} =  -\sum_{j=1}^{\infty} \frac{\left( (\partial \hat{\phi}^{(0)}  )^2+H^2 (\hat{\phi}^{(0)})^2 \right)^{j}}{H^{2j}} f_j\left( \hat{\phi}^{(0)} \right),
\label{eq:eqwithfjs}
\ee
where $\hat{\phi}^{(i)} \equiv -\alpha H^2 \phi^{(i)}/\Lambda^{(D+2)/2}$. The first terms of the power series $f_j$ are given in Table~\ref{tab:functions}.

\section{Broken diffeomorphisms and Goldstone modes}
\label{sec:broken-diffs}

In this section, we outline a geometric interpretation of all the dS scalar field theories with nonlinear symmetries---there is a unified description of them in terms of spontaneously broken diffeomorphisms. This discussion parallels that of Ref.~\cite{Roest:2020vny} for the special galileon in flat space.

\subsection{Nonlinearly realized extensions of the isometry algebra}
Under a general infinitesimal coordinate transformation $\delta x^\mu = - \xi^\mu(x)$, the metric transforms as 
 \be
  \delta {g}_{\mu\nu} = -2 {\nabla}_{(\mu} \xi_{\nu)}  \,. \label{diff-transf}
 \ee
A special class of diffeomorphisms are those that leave the metric invariant. These correspond to Killing vectors, which generate the isometries of the space. For example,
in stereographic coordinates, $X^{\mu}(x)=\frac{x^{\mu}}{1+H^2 x^2/4}$ and $X^{D+1}(x)=\frac{1}{H}\frac{1-H^2 x^2/4}{1+H^2 x^2/4}$, the dS metric is
 \begin{align}
\rd s^2 = \frac{1}{\left(1 + \frac{H^2 }{4}x^2 \right)^2} \eta_{\mu \nu} \rd x^\mu \rd x^\nu \,,
 \end{align}
where $\eta= {\rm diag} (-1, 1, \dots, 1)$ and  $x^2 \equiv \eta_{\mu\nu} x^\mu x^\nu$, and the Killing vectors are given by 
\be
  \xi^\mu = \omega^\mu{}_\nu x^\nu + \epsilon^\nu\left[\delta_\nu^\mu\left(1 - \frac{H^2}{4}x^2\right)  + \frac{H^2}{2} x^\mu x_\nu\right] \,,
\ee
for constants $\epsilon^\mu$, $\omega^\mu{}_\nu$, where 
indices are raised and lowered with $\eta_{\mu\nu}$
and $\eta_{\mu \lambda} \omega^\lambda{}_\nu=-\eta_{\nu \lambda} \omega^\lambda{}_\mu$.

Any two vector fields commute into a third via the Lie bracket, 
 \begin{align}
  \xi_3^\mu = \xi_1^\nu \nabla_\nu \xi_2^\mu -  \xi_2^\nu \nabla_\nu \xi_1^\mu \,.
 \end{align}
 The total set of vector fields forms an infinite-dimensional Lie algebra with the Lie bracket as the commutator.
The commutator of two Killing vectors gives another Killing vector, so the isometries form a finite-dimensional Lie subalgebra of the algebra of all vector fields.
Our goal is to find an intermediate-sized algebra containing the subalgebra of isometries by including additional diffeomorphisms, which will now be nonlinearly realized on a scalar Goldstone mode instead of being isometries of the space. Importantly, we still wish the algebra to be finite dimensional.
Since we want the Goldstone mode to be a scalar field, we consider diffeomorphisms whose generators are exact as one forms, i.e., $\xi_\mu = \nabla_\mu \sigma$.  The dS Killing vectors are not exact, as can be verified by showing that they are not closed.
 
Taking the commutator of an exact diffeomorphism $\nabla_\mu\sigma$ and a Killing vector $\xi^{\mu}$ gives another exact diffeomorphism $\nabla_\mu\sigma'$ with $\sigma' \equiv \xi^\mu \nabla_\mu \sigma$, so the exact diffeomorphisms form a linear representation of the isometry group.

Commuting two exact diffeomorphisms formed from $\sigma_a$ and $\sigma_b$ gives a diffeomorphism generated by the vector 
\be
\xi_{ab}^{\mu} \equiv \nabla^{\nu} \sigma_a \nabla_{\nu} \nabla^{\mu} \sigma_b-\nabla^{\nu} \sigma_b \nabla_{\nu} \nabla^{\mu} \sigma_a.
\ee
One way to close the algebra is to require that $\xi_{ab}^{\mu}$ is itself a Killing vector, which gives the condition
 \begin{align}
  \nabla_{(\mu} \xi^{ab}_{\nu)} = \nabla^\rho \sigma_a \nabla_{(\mu} \nabla_{\nu} \nabla_{\rho)} \sigma_b -  \nabla^\rho \sigma_b \nabla_{(\mu} \nabla_{\nu} \nabla_{\rho)} \sigma_a \, = 0.
 \end{align}
 We do not attempt to solve this equation in general, but rather restrict to the following simpler sufficient condition that will be enough to encompass the exceptional scalar theories of interest:
 \begin{align}
 \nabla_{(\mu} \nabla_\nu \nabla_{\rho)} \sigma_a =\lambda \, g_{(\mu \nu} \nabla_{\rho)} \sigma_a \,, \quad\qquad \nabla_{(\mu} \nabla_\nu \nabla_{\rho)} \sigma_b=\lambda\, g_{(\mu \nu} \nabla_{\rho)} \sigma_b\,, \label{condition}
 \end{align}
 where $\lambda$ is some constant.

Before discussing the possible solutions to Eq.~\eqref{condition}, let us see the relevance for the nonlinear scalar symmetries. The algebra including exact diffeomorphisms can be realized nonlinearly on a Goldstone mode $\phi$, provided the Goldstone mode transforms in the usual linear way under the Killing vectors and nonlinearly under the exact diffeomorphisms according to
 \begin{align} 
  \delta_{\sigma_a} \phi = \sigma_a + \frac{\alpha}{\Lambda^{(D+2)/2}} \nabla^\mu \sigma_a \nabla_\mu \phi \,.\label{phitransphie}
 \end{align} 
 For example, the commutator of two of these transformations is
 \be
 [\delta_{\sigma_a}, \delta_{\sigma_b}] \phi = -\frac{\alpha^2}{\Lambda^{D+2}} \xi^{\mu}_{ab} \nabla_{\mu} \phi.
 \ee
 From $\phi$  and the dS metric $g_{\mu\nu}$, we can then build a metric that transforms covariantly under the extended algebra, which can then be used to construct invariant interactions. 

Returning to the condition \eqref{condition}, we can identify the following three sets of solutions that separately form irreducible representations of the isometry group:
 \begin{itemize}
 \item
 $\boldsymbol{k=0}$: This case corresponds to the trivial solution with $\nabla_\mu \sigma = 0$, i.e., $\sigma$ is constant.  This is a degenerate case because the parameter $\sigma$ corresponding to an exact diffeomorphism $\nabla_\mu\sigma$ is only defined up to a constant, though the transformation \eqref{phitransphie} depends on this constant.    This constant represents a simple shift symmetry of the scalar.  From this point of view, the shift symmetry is not really an extension of the spacetime isometries, but instead a kind of central extension realized by passing to the nonlinear realization \eqref{phitransphie}.
 \item
$\boldsymbol{ k=1}$: This case consists of the solutions with $\nabla_{(\mu} \nabla_{\nu)} \sigma_a = -H^2 g_{\mu \nu} \sigma_a$. This implies that the exact diffeomorphisms are conformal Killing vectors. There are $D+1$  conformal Killing vectors, corresponding to dilations and special conformal transformations. In stereographic coordinates, the solutions read
 \begin{align}
    \sigma = \frac{1 - x^2 H^2/4 }{1+x^2H^2/4} ( c+c_\mu y^\mu) \,, \qquad y^\mu \equiv x^\mu / (1 - x^2 H^2/ 4) \,.\label{k1solepe}
 \end{align}
These can be written as the restrictions of the ambient space objects $S_A X^A$. 
The Killing and conformal Killing vectors together generate the algebra $\mathfrak{so}(D,2)$. The associated Goldstone mode is the dilaton. The following metric transforms covariantly under the nonlinearly realized symmetries 
\be
G^{(1)}_{\mu\nu} \equiv e^{2\hat{\phi}}g_{\mu\nu}, \quad \hat{\phi} \equiv -\frac{\alpha H^2}{\Lambda^{(D+2)/2}}\phi , 
\ee
and can be used to construct invariant Lagrangians as diffeomorphism invariants (aside from a single Wess--Zumino term)~\cite{Hinterbichler:2012mv}.
\item
$\boldsymbol{k=2}$: Lastly, we have the solutions with $ \nabla_{(\mu} \nabla_\nu \nabla_{\rho)} \sigma_a =-4 H^2 g_{(\mu \nu} \nabla_{\rho)} \sigma_a $. These read in stereographic coordinates
  \begin{align}
   \sigma =  \left( \frac{1 - x^2H^2 / 4}{1+x^2 H^2/4} \right)^2 ( c + c_\mu y^\mu + c_{\mu \nu} y^\mu y^\nu)  \,. \label{quadratic}
 \end{align}
  These can be written as the restrictions of the ambient space objects $S_{AB} X^AX^B$. 
The associated Goldstone mode will be the special galileon with the nonlinearly realized algebra $\mathfrak{sl}(D+1, \mathbb{R})$. In this case, we can construct the metric \eqref{eq:k2metric},
\be
G_{\mu\nu}^{(2)} \equiv g_{\mu\nu}-H^{-2}\nabla_\mu\nabla_\nu \hat{\phi}+H^{-2}\nabla_\mu \hat{\phi} \nabla_\nu \hat{\phi},  \quad \hat{\phi} \equiv -\frac{2 \alpha H^2}{\Lambda^{(D+2)/2}}\phi , 
\ee
which transforms covariantly under the nonlinearly realized symmetries.\footnote{Actually, it transforms covariantly under a slightly larger set of symmetries, namely $\mathfrak{gl}(D+1, \mathbb{R})$. However, since the special galileon Lagrangian is a Wess--Zumino term, it cannot be written in terms of this metric, so it only has $\mathfrak{sl}(D+1, \mathbb{R})$ symmetry. Similarly, the appearance of $e^{-(D+1) \hat{\phi}}$ in Eq.~\eqref{eq:sgalEOM} breaks the symmetry of the equation of motion down to $\mathfrak{sl}(D+1, \mathbb{R})$.}
\end{itemize}

All of the above solutions can be written in the form
 \begin{align}
 \sigma =  \left( \frac{1 - x^2 H^2 / 4}{1+x^2H^2/4} \right)^k P_k (y) \,, \label{sigma}
 \end{align}
where $k \in \{0,1, 2\}$ and $P_k$ are $k$-th order polynomials in $y$, as given in Eqs.~\eqref{k1solepe} and \eqref{quadratic} for $k=1, 2$.  These solutions can be lifted to the ambient space functions
\be
\Sigma = \frac{1}{(X^2 H^2)^{k/2}}S_{A_1 \dots A_k} X^{A_1} \dots X^{A_k},
\ee
where the overall factor of $(X^2H^2)^{-k/2}$ is chosen so that $\Sigma$ has weight zero. The nonlinear realization \eqref{phitransphie} of the exact diffeomorphisms on the Goldstone mode $\phi$ can then be written in ambient space as
\begin{align}
\label{eq:goldstone-ambient}
\delta \Phi = \Sigma +  \frac{\alpha H^2 X^2}{ \Lambda^{(D+2)/2}} \nabla_A \Sigma \nabla^A \Phi \,,
\end{align}
where the ambient space field $\Phi$ has weight zero. 

This discussion shows how we can identify the transformations of Goldstone scalars in dS space with various subsets of non-Killing diffeomorphisms.  The trivial case $k=0$ yields a scalar field with a constant shift symmetry. The cases $k=1$ and $k=2$ are the curved-space analogues of the flat space dilaton and special galileon. For flat space, it is known that the corresponding algebras $\frak{so}(D, 2)$ and $\frak{igl}(D)=\frak{gl}(D,\mathbb{R}) \ltimes \mathbb{R}^D$ are the only finite-dimensional Lie subalgebras of the diffeomorphism algebra properly containing the Poincar\'e isometry algebra \cite{xthschool}.\footnote{Note that flat space DBI has as its symmetry algebra $\frak{iso}(D,1)$, which is not a subalgebra of $D$-dimensional diffeomorphisms, so it is not realized in this way.} 
We do not know if a similar statement has been proven for dS space.

\section{Conclusions}

We have elucidated the structure of exceptional effective field theories of a single self-interacting scalar field on dS space.  A particularly interesting example is the $k=2$ theory with a mass $m^2=-2(D+1)H^2$, which is the dS version of the special galileon.  We found a simpler formulation of this theory and showed that it is related to the original formulation of Ref.~\cite{Bonifacio:2018zex} by a field transformation that can be thought of as a dS uplift of galileon duality.  The $k=1$ theory, with a mass $m^2=-D\,H^2$, has two natural formulations: as the DBI theory of a dS$_D$ brane probing an AdS$_{D+1}$ bulk and as the dilaton of broken conformal symmetry on dS space.  We found strong evidence that these two formulations are also equivalent, related by a field transformation that is a dS uplift of the transformation of Ref.~\cite{Bellucci:2002ji, Creminelli:2013fxa}.  Finally, we gave a new interpretation of these symmetries as sets of broken diffeomorphisms.

The coset construction gives a systematic way to build Goldstone effective field theories that are invariant under a given symmetry breaking pattern for internal symmetries \cite{Coleman:1969sm, Callan:1969sn,Volkov:1973vd}.  The effective field theories obtained are unique, in the sense that any two theories with the same breaking pattern will be equivalent under field redefinitions of the Goldstones.  The Goldstones parametrize the coset, and the field redefinitions are different parameterizations of the coset.   For spacetime symmetries, the coset construction is more involved \cite{Volkov:1973vd,xthschool}, and the number of the Goldstones is not generally equal to the dimension of the coset.  As far as we know, there is no proof that effective field theories with the same spontaneous spacetime symmetry breaking pattern, and the same degrees of freedom, are always equivalent (this is expected to be the case, although there are subtleties when coupling to matter \cite{Creminelli:2014zxa,Dubovsky:2007ac}).  Our results here give more evidence that they are in fact the same; different realizations of the same symmetry breaking pattern on the fields give theories whose Lagrangians look very different, but which are related by non-trivial field redefinitions involving all orders in powers of the field.

A natural extension of our study would be to explore where there are similarly simple (A)dS formulations of shift-symmetric theories that involve higher-spin particles.  For example, there are shift-symmetric theories containing spin-1 fields on curved and flat space \cite{DeRham:2018axr,Bonifacio:2019hrj, Kampf:2021bet} that may have more unified formulations, and there may be higher-spin examples as well \cite{Bonifacio:2018zex}.

An important open question is to understand the on-shell avatar of the symmetries of these exceptional scalar theories in curved spacetimes, where the natural observable quantities are boundary correlation functions. This is the purview of the cosmological bootstrap, which aims to classify the space of possible cosmological correlators, and so constrain the physics of inflation (see, e.g., Refs.~\cite{Arkani-Hamed:2018kmz,Sleight:2019hfp,Baumann:2019oyu,Baumann:2020dch,Pajer:2020wxk,Meltzer:2020qbr,Goodhew:2020hob,Melville:2021lst,Jazayeri:2021fvk,Baumann:2021fxj,Bonifacio:2021azc,Meltzer:2021zin,Sleight:2021plv,Hogervorst:2021uvp,DiPietro:2021sjt} for recent developments). It is natural to wonder about the role of these scalar theories in the bootstrap construction: how precisely are their special properties  reflected in their (A)dS correlators? We expect that, much like in the flat space case, their correlation functions will display a rich web of interconnections, which could help reveal some underlying general structures of cosmological correlators. 
An intriguing direction to pursue is to search for (A)dS versions of the double copy and scattering equations. Some aspects of color-kinematics duality and the double copy have been studied in Refs.~\cite{Farrow:2018yni,Armstrong:2020woi,Albayrak:2020fyp,Alday:2021odx,Zhou:2021gnu,Sivaramakrishnan:2021srm}, but much of the structure remains mysterious. Similarly, scattering equations in AdS have  just begun to be explored~\cite{Eberhardt:2020ewh,Roehrig:2020kck,Gomez:2021qfd}. It is natural to expect that these exceptional scalar theories will be similarly helpful in elucidating these structures as they were in flat space.

\paragraph{Acknowledgements:} We would like to thank Tanguy Grall, Hayden Lee, and David Stefanyszyn for helpful discussions. During this work we made use of the \texttt{Mathematica} package \texttt{xAct} \cite{xAct}. JB is supported by the research program VIDI with Project No.~680-47-535, which is partly financed by the Netherlands Organisation for Scientific Research (NWO), and partially supported by STFC HEP consolidated grant ST/T000694/1.  KH acknowledges support from DOE grant DE-SC0009946 and from Simons Foundation Award Number 658908.

\renewcommand{\em}{}
\bibliographystyle{utphys}
\addcontentsline{toc}{section}{References}
\bibliography{AdSscalars}

\end{document}